\documentclass[a4paper,11pt]{article}
\pdfoutput=1 

\usepackage{jinstpub} 
\usepackage[normalem]{ulem}
\usepackage{gensymb}
\usepackage[labelformat=simple]{subfig}
\usepackage{caption}

\graphicspath{{Figures/}{./}}
\title{A simulation study to distinguish prompt photon from $\pi^0$ and beam halo in a granular calorimeter using deep networks}

\author[a]{S. Ghosh,}
\author[b,c,1]{A. Harilal, \note{Corresponding author.}}
\author[b,d]{A. R. Sahasransu,}
\author[b]{R. K. Singh}
\author[a]{and S. Bhattacharya}

\affiliation[a]{High Energy Nuclear and Particle Physics Division, Saha Institute of Nuclear Physics, HBNI,\\ 1/AF Bidhannagar, Kolkata, India}
\affiliation[b]{Department of Physical Sciences, Indian Institute of Science
Education and Research Kolkata,\\ Mohanpur, 741246, India}
\affiliation[c]{Department of Physics, Carnegie Mellon University, Pittsburgh, USA}
\affiliation[d]{Vrije Universiteit Brussel, Belgium}

\emailAdd{aharilal@andrew.cmu.edu}

\abstract{In a hadron collider environment identification of prompt photons originating in a hard partonic scattering process and rejection of non-prompt photons coming from  hadronic jets or from beam related sources, is the first step for study of processes with photons in final state. Photons coming from decay of $\pi_0$'s produced inside a hadronic jet and photons produced in catastrophic bremsstrahlung by beam halo muons are two major sources of non-prompt photons. 
In this paper the potential of deep learning methods for separating the prompt photons from beam halo and $\pi^0$'s in the electromagnetic calorimeter of a collider detector is investigated, using an approximate description of the CMS detector. 
It is shown that, using only calorimetric information as images with a Convolutional Neural Network, beam halo (and $\pi^{0}$) can be separated from photon with 99.96\% (97.7\%) background rejection for 99.00\% (90.0\%) signal efficiency which is much better than traditionally employed variables.
}


\begin{document}
\maketitle
\flushbottom

\section{Introduction}
\label{S:1}


Understanding the symmetry breaking mechanism in the standard model (SM) and search for new physics beyond the standard model are major goals of the physics program of the Large Hadron Collider (LHC).  Photons and electrons play important roles in these searches as they provide clean signatures in hadron environments. Distinguishing prompt photons (coming from hard scattering) from photons coming from neutral meson ($\pi^0,\eta$) decays or other anomalous sources (e.g. bremsstrahlung photon coming from beam halo muon) is of critical importance. An example is the search for  the Higgs boson using its H$\rightarrow  \gamma\gamma$ decay mode, which was one of the main channels for discovery of Higgs boson. The di-photon final state produced by this decay, has a large background coming from hadronic jets, where most of the jet energy has gone to single $\pi^0$. Sensitivity of this search depends largely on the power of rejection of this background. Another example is search for dark matter and large extra dimensions in final states with a photon and large missing transverse momentum. In this channel, along with the photons coming from pion decays,  beam halo photons pose an additional challenge. \\
Currently employed traditional techniques use combination of shower shape variables which are intelligently constructed by the physicist, to try to capture the difference in spatial patterns of the signal and the background events, from the energy deposit per crystal information of an electromagnetic calorimeter, but are only a few in a set of infinite such possible variables. Classification using supervised learning algorithms like an Artificial Neural Network (ANN) or a boosted decision tree (BDT), with the high level features made by physicists as inputs, have also been done successfully~\cite{Khachatryan:2014ira}. A prime example of such an analysis is the search for Higgs boson in the Higgs to diphoton channel performed by the CMS experiment at the LHC, where a boosted decision tree was deployed for identification of prompt photons~\cite{HggDisco:2012hg}. However, with the advent of the the new image recognition networks it should now be possible to benefit from these modern techniques.
Only recently significant efforts have been made in this direction. 
There are a number of recent instances in High Energy Physics where physics classification problems have been recast as computer vision problems like neutrino classification~\cite{Aurisano:2016jvx, Acciarri:2016ryt} and jet image classification~\cite{deOliveira:2015xxd, Baldi:2016fql,  Barnard:2016qma, Komiske:2016rsd}.\\
The analysis presented in this paper derives its motivation from such recent instances of using the emerging techniques in Deep Learning (DL) like Convolutional Neural Network (CNN)~\cite{lecun:98}. In this approach the machine learns to construct many high level feature variables starting from the energy deposit per crystal information of an electromagnetic calorimeter. Every filter in the CNN projects such a high level variable.
A CNN is thereby employed in trying to extract maximum amount of information from the raw output from the electromagnetic calorimeter, which can lead to better performance in discriminating between the two classes. In this study, data directly from the detector as images are used and a CNN is run on them without providing any high level physics information. Performance of the CNN is compared with a Multi Layer Perceptron (MLP) fed with specialized physics variables so as to evaluate the efficacy of these deep learning algorithms in identifying features from the data without much human input. \\
The rest of the paper is outlined as follows: in Section~\ref{S:2}, an outline of photon identification in high energy physics and the various classes of photons considered in this paper are given. In Section~\ref{S:3}, the details of the detector simulation are described, and in Section~\ref{S:4}, the full analysis and results are presented, followed by a conclusion of this study in Section~\ref{S:5}.

\section{Photon Identification}
\label{S:2}

\subsection{Prompt photons}
\begin{figure}[htbp]
\subfloat[(a)]{\includegraphics[width=0.5\textwidth]{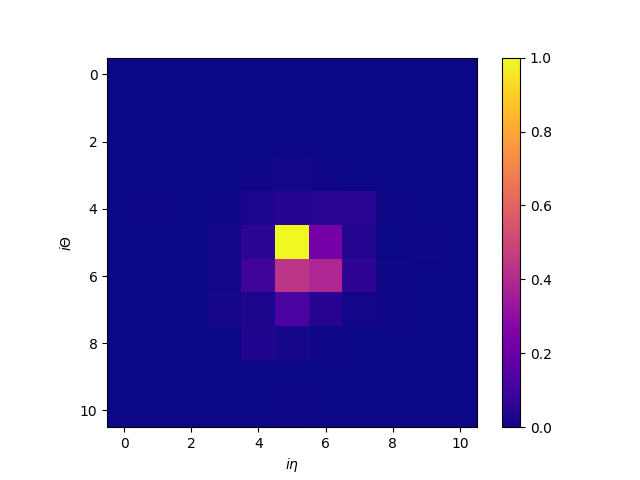}}
\hspace{0.3cm}
\subfloat[(b)]{\includegraphics[width=0.5\textwidth]{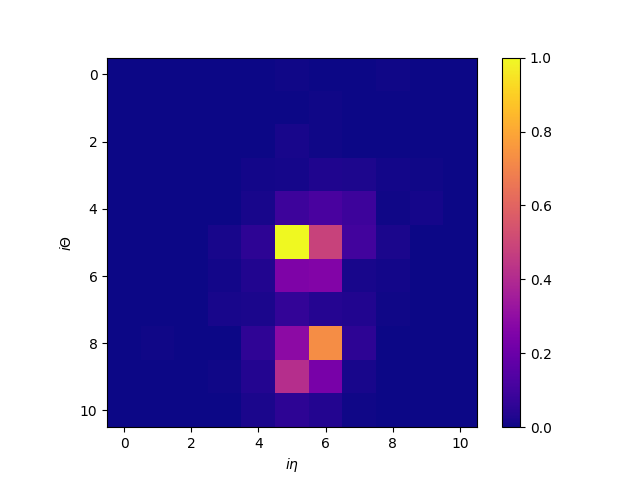}}
\vfill
\subfloat[(c)]{\includegraphics[width=0.5\textwidth]{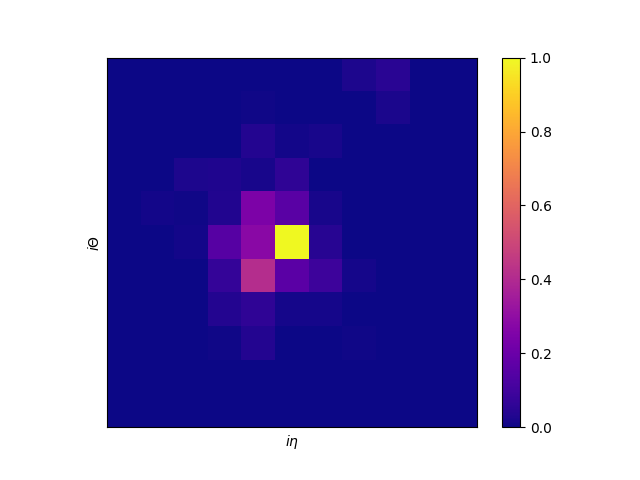}}
\subfloat[(d)]{\includegraphics[width=0.5\textwidth]{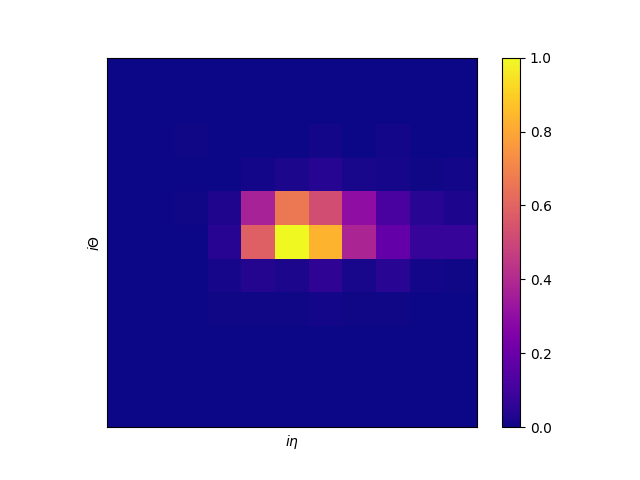}}
\caption{The image representing the energy deposit pattern of (a) a prompt photon, (b) a converted photon, (c) a $\pi^0$, (d) a beam halo photon energy deposit map in the Ecal, normalized with the seed crystal energy in $\eta, \phi$ coordinates. The colors represent the amount of energy deposited in a crystal.}
\label{fig:all_photon}
\end{figure}

Prompt photons are photons produced in $qg$ Compton scattering, $q\bar{q}$ annihilation or in $gg$ fusion. 
The tracker material in front of the electromagnetic calorimeter (ECAL) causes photons to convert to $e^+ e^-$ pairs in the tracker before reaching the ECAL. Unlike the photons that pass through the tracker unconverted, these converted photons have hits in the tracker and have ECAL energy deposition more spread in $\phi$ due to bending of the $e^+ e^-$ trajectories in the strong magnetic field. 
Typical energy deposit maps of different classes of photons are shown in Figure~\ref{fig:all_photon}.

\subsection{Photons from decay and fragmentation}
\begin{figure}[htbp]
\subfloat[(a)]{\includegraphics[height=4cm, width=0.3\textwidth]{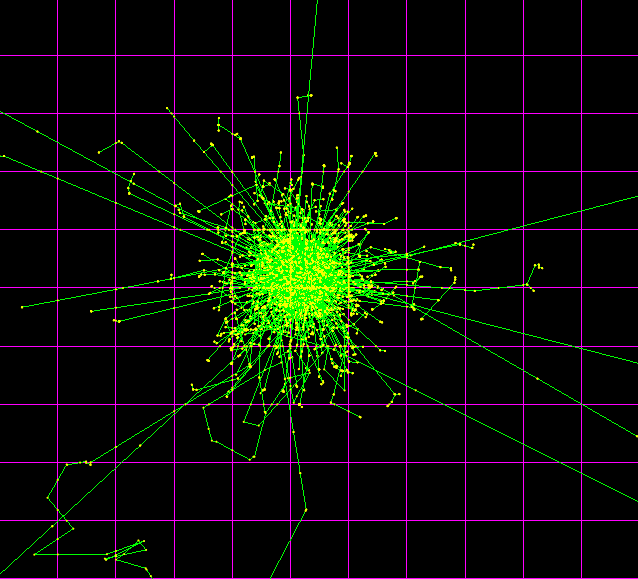}}
\hspace{0.3cm}
\subfloat[(b)]{\includegraphics[height=4cm, width=0.3\textwidth]{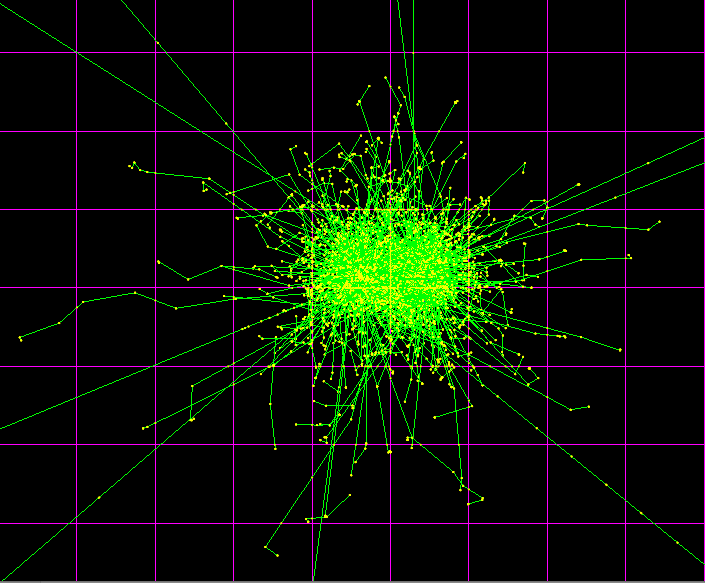}}
\hspace{0.3cm}
\subfloat[(c)]{\includegraphics[height=4cm, width=0.3\textwidth]{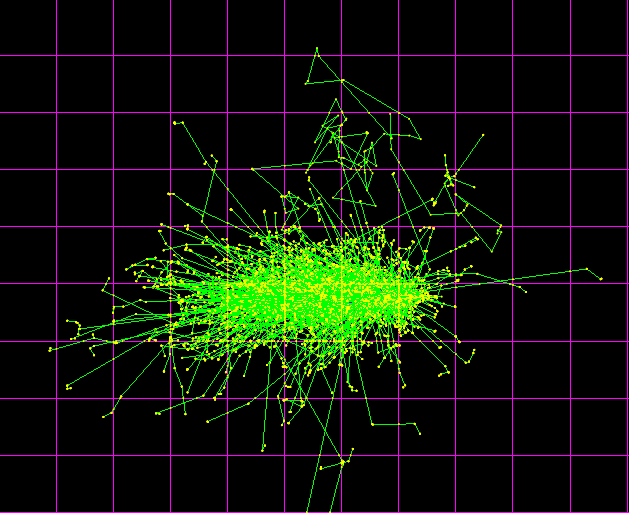}}
\caption{Showers from a 10 GeV (a) prompt photon, (b) $\pi^0$ and (c) beam halo photon. For (a) and (b), photons were shot perpendicular to the surface of the calorimeter (in this case perpendicular to the plane of the paper) and for (c), photon was shot from the side (in this case from the right side of the paper).}
\label{fig:sim}
\end{figure}

Photons from decay of hadrons such as $\pi^0, \eta$, fragmentation process as well as photons from bremsstrahlung of charged particles faking a prompt photon, form a large background over the signal. Diphoton decay of a neutral meson ($\pi^0, \eta$) inside a hadronic jet is the single largest background to prompt photons. In this study, the two photons from a $\pi^0$ above 10 GeV hit the same crystal in the ECAL and appear like one photon. The hadronic multijet production cross section at the LHC is orders of magnitude higher than a typical new physics process. These jets copiously produce neutral mesons, a small fraction of which can fake a prompt photon. For $H \rightarrow \gamma \gamma$  analysis this background is almost two orders of magnitude larger than the signal, assuming a jet faking photon rate of \textasciitilde  $0.1\%$~\cite{Pieri:2006bm}.\\\\
The energy deposits of prompt photons, and photons from $\pi^0$ and beamhalo in the ECAL as seen from the interaction point of our simulation is shown in Figure~\ref{fig:sim}.


\subsection{Beam halo photons}
One of the main backgrounds, known as machine induced background (MIB), in high energy particle detectors comes from particles entering the detector from the accelerator. These particles which are produced in the hadronic and electromagnetic showers resulting from beam protons interacting with collimators or residual gas molecules in the vacuum pipe are called beam halo. Pions, being the lightest hadrons, are produced easily in these hadronic interactions, and constitute majority of the beam halo. Being short lived, the neutral pion decays into two photons, and the charged pions decay into muons. Some of these muons are very energetic with energies of hundreds of GeV, i.e, greater than the critical energy of muons in the lead tungstate crystals of the calorimeter. These high energy muons undergo bremsstrahlung as they interact with the atoms of the crystals, and emit photons.
These halo photons result in final states with a single photon with no other object to balance its energy and momentum, thus mistakenly pointing to final states with invisible particles recoiling against photons. These photons are along the direction parallel to the beam pipe, and so have an elongated shower along $\eta$. 
\section{Simulation}
\label{S:3}
\begin{figure}[ht!]
\centering
\includegraphics[width=1.1\textwidth]{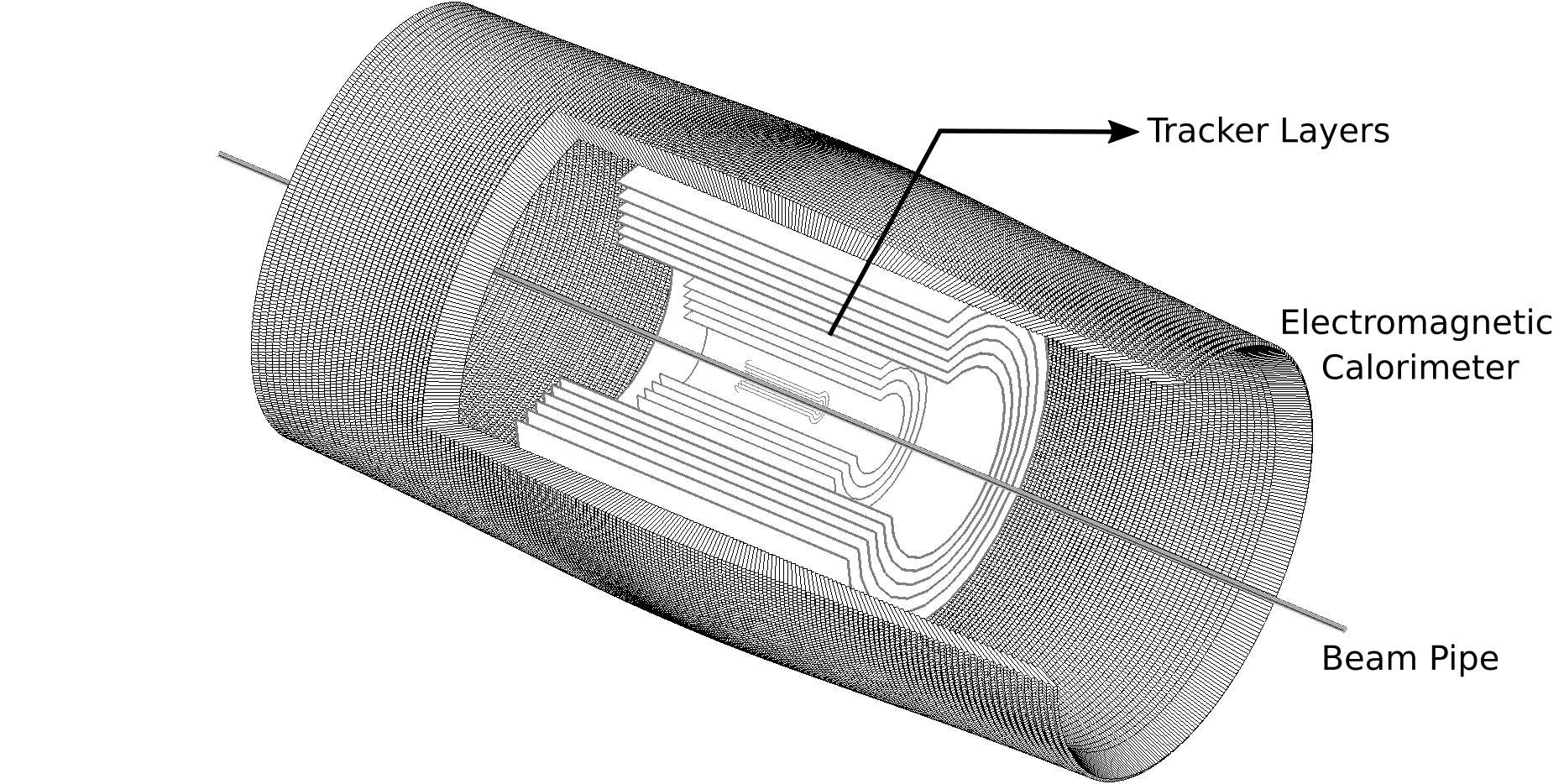}
\caption{The Cylindrical detector simulated in \texttt{GEANT4}.}
\label{fig:cyl_geoml}
\end{figure}

A model detector comprising of calorimeter and tracker has been constructed~\cite{github:2018} using \texttt{GEANT4}~\cite{Agostinelli:2002hh} to resemble CMS from its technical design report (TDR)~\cite{cmstdr:2000eqx}. The calorimeter construction includes the barrel region between pseudorapidity $-1.479< \eta < 1.479$. It is made of parameterized volumes of part of a sphere arranged as an array of $ \mathrm{PbWO_4}$ crystals placed in a cylindrical arrangement. Each parameterized volume has $0.0174$ $\eta$ coverage, $1 \degree $ azimuthal angle ($\phi$) coverage and $22\ \rm{cm}$ radial length. The size of the crystals is $2.2\ \rm{cm} \times 2.3\ \rm{cm} $ in the front face and vary from $2.4\ \rm{cm} \times 2.4\ \rm{cm} $ to $ 2.6\ \rm{cm}\times 2.7\ \rm{cm} $ in the back face. These values are within $5 \%$ of the ECAL crystal sizes in the TDR. A greatly simplified version of a tracker has been implemented as $13$ concentric cylinders of varying thickness of silicon. The first three layers have a thickness of $285 \rm{\mu m}$ followed by four layers of $320\ \rm{\mu m}$ and six more layers of $500\ \rm{\mu m}$ $\rm{Si}$ thickness. Additional material has been added to have a material budget similar to that of the CMS tracker in the pseudo-rapidity region $-0.8 \le \eta \le 0.8$. A  uniform magnetic field of 4 T has been applied along the positive z-axis. 
The cross-sectional view of the geometry is shown in Figure~\ref{fig:cyl_geoml}.
\\
The simulation has a tag for filtering out photons which convert anywhere inside the tracking volume.
The standard physics list FTFP-BERT has been used for simulating physics process to keep the simulation as close to the general purpose detectors CMS and ATLAS as possible~\cite{FTFP_in_EHEP}. These include bremsstrahlung, pair production and photo-electric effect for photons, as well as Compton scattering for $e^-$. The particle is assumed to deposit its entire energy at a point if it can't travel a distance of $1$ mm further from the point.\\
\section{Analysis and Results}
\label{S:4}
This section describes the details of the analysis procedure and results. The information from the cylindrical ECAL has been represented as a 2D image in the $\eta \times \phi$ space as n$\times$n (n = 11 or 25 depending on the problem described in sections below) matrix of cells around a local maximum in energy deposit (seed crystal). The cells represent calorimeter crystals and the values of the cells the energy contained in them. The values have been further normalized to the seed crystal energy before feeding the matrix of cells as an input to the networks. Three different network analyses have been performed for each classification problem : 
\\
\begin{itemize}
    \item The traditional shape variables constructed out of cell values fed to an MLP or DNN.
    \item The normalized cell values fed to an MLP or DNN.
    \item n$\times$n matrix of normalized cell energies fed to a CNN.
\end{itemize}

In the first analysis the shape variables are constructed from intuition, utilizing the knowledge about the narrow lateral shower profile of an electromagnetic shower. Some variables used are designed after the standard shower shapes variables used in the CMS ECAL~\cite{CMS-PAS-EGM-10-005, CMS-PAS-EGM-10-006} and other prior studies done on homogeneous, granular calorimeter hodoscopes.

The main study with the cylindrical geometry described in section~\ref{S:3}, has been cross checked with a planar geometry with a magnetic field parallel to the face of the crystals. 
\subsection{Network Architecture and Ranking}
\begin{figure}[ht!]
\centering
\includegraphics[width=1.\textwidth]{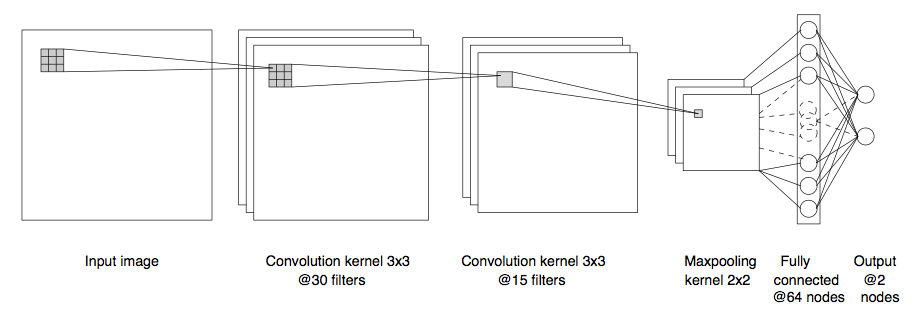}
\caption{The CNN architecture. The input image is of size 11$\times$11 for beam halo rejection and 25$\times$25 for $\pi^0$ rejection. }
\label{fig:CNN_model}
\end{figure}

\begin{figure}[ht!]
\centering
\includegraphics[width=0.5\textwidth]{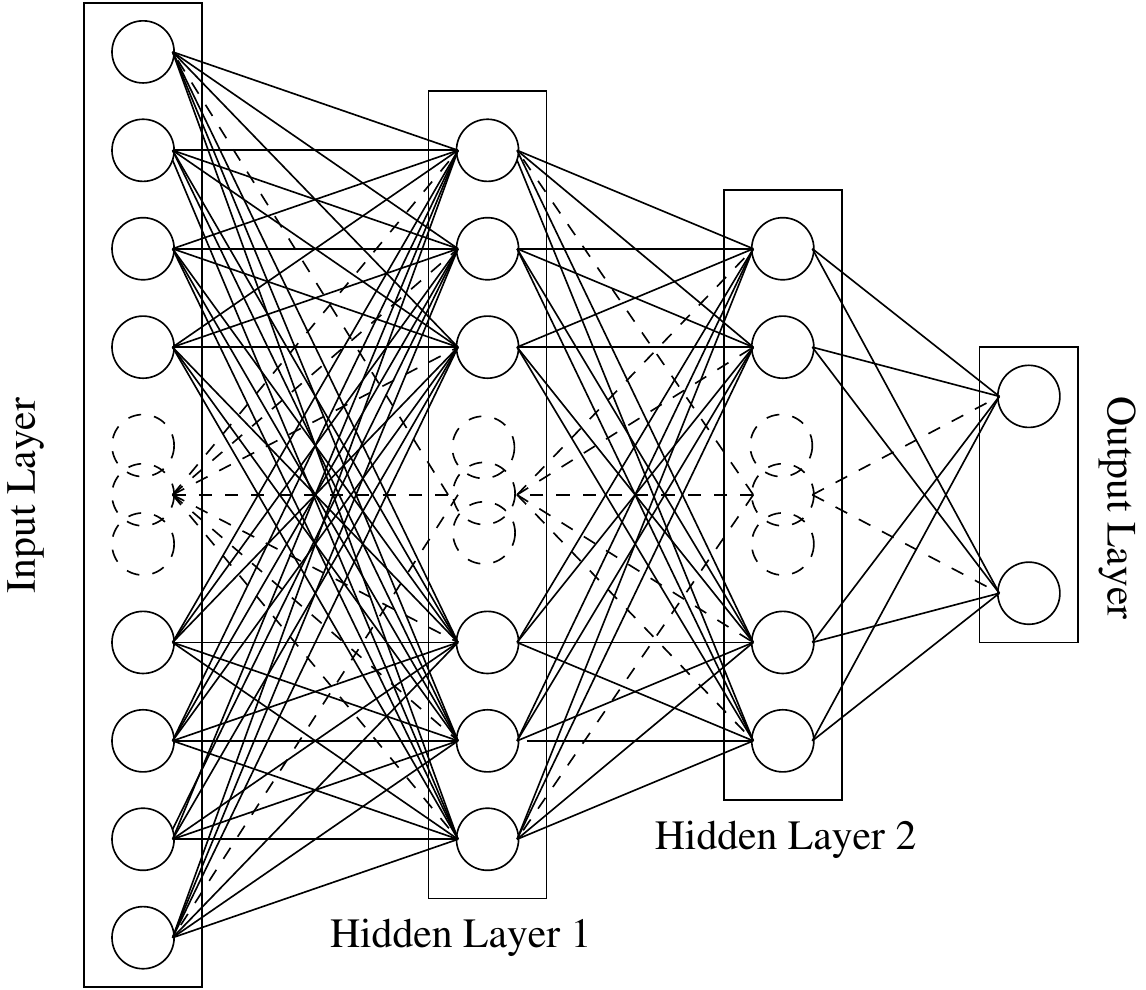}
\caption{The network for $\pi^0$ separation (DNN) has two hidden layers with 64 and 32 nodes respectively, with RELU activation. For the beam halo separation with MLP , the network has only one hidden layer of 64 nodes with RELU activation. The output layer is acted on by softmax activation.  }
\label{fig:ANN_model}
\end{figure}

The package \texttt{Keras}~\cite{site:keras}, with \texttt{Tensorflow}~\cite{site:tf} as backend has been used for implementing the networks used in the analysis. The various networks used are listed below:

\begin{description}
\item [Convolutional Neural Network (CNN):] A CNN has been constructed with two convolutional layers with filters of size 3$\times$3 and stride 1$\times$1, acted on with activation function of rectified linear unit (RELU)~\cite{Nair:2010:RLU:3104322.3104425} on the outputs along with L2 regularization on the network weights, followed by maxpooling of pool window size 2$\times$2. It is followed by a fully connected layer of 64 nodes, with dropout regularization~\cite{JMLR:v15:srivastava14a} of 30\%. Finally there is a fully connected layer with the softmax activation function giving a binary output.
The structure of the CNN used is shown in Figure~\ref{fig:CNN_model}.
\item [Multi Layer Perceptron (MLP):] For the photon-beam halo separation, a MLP with one hidden layer of 32 nodes with RELU activation, and an output layer of 2 nodes with softmax activation is used. We see that optimal classification is obtained using this simple  network. The network architecture is shown in Figure~\ref{fig:ANN_model}, where only one hidden layer is considered. 
\item [Deep Neural Network (DNN):] For photon-$\pi^0$ separation, a MLP with two hidden layers is used, as the problem is more difficult and a deeper network is found to perform better. The first layer has 64 nodes, the second layer has 32 nodes, both with RELU activation, followed by a dropout of 30\%, and the output layer has 2 nodes with softmax activation. The structure of the MLP used is shown in Figure~\ref{fig:ANN_model}. 

\end{description}
All the networks use cross-entropy loss function with the ADADELTA optimizer~\cite{DBLP:adadelta}.\\
Receiver Operating Characteristics (ROC) curve with signal efficiency vs. background rejection have been plotted to evaluate the performance of each classifier. The area under the curve (AUC) of the ROC has also been used as a quality criterion for the classification.
\subsection{Datasets}
51,000 images each of the signal (prompt photons) and background (non-prompt photons) classes have been generated. For prompt photon - beam halo separation, images of dimensions 11$\times$11 have been used, while for prompt photon - $\pi^0$ separation, images of dimensions 25$\times$25 have been used to efficiently capture the conversion of photons. 
Out of the total sample set, 80\% has been used for training and 20\% for testing. Out of the training set, 30\% has been used for validation. Training was stopped when the loss on the validation step did not decrease significantly after a certain number of epochs.

\subsection{Beam halo - prompt photon separation}
\label{S:halo}
\begin{figure}[h]

\subfloat[(a)]{\includegraphics[height=4.0cm,width=0.35\textwidth]{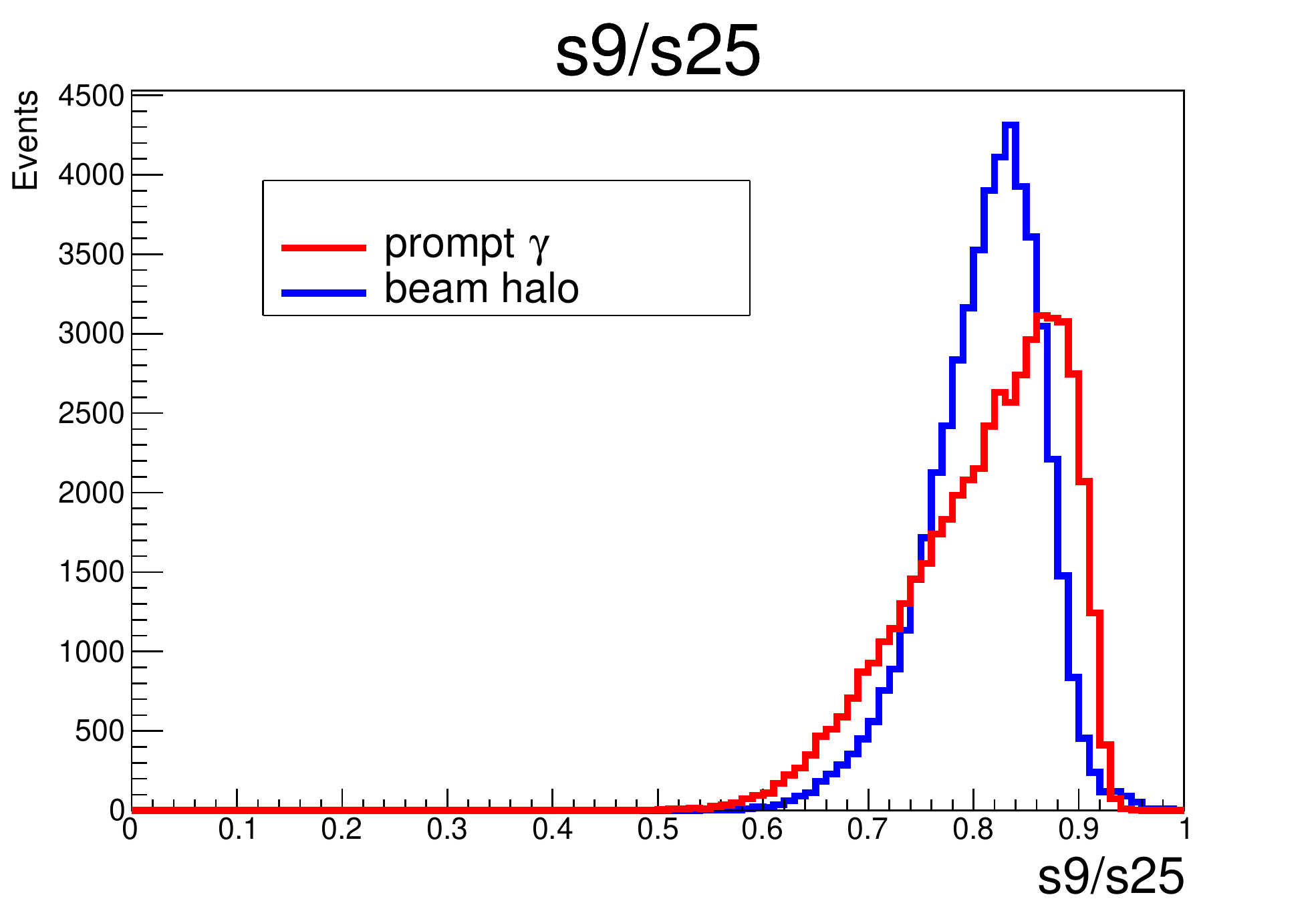}}
\subfloat[(b)]{\includegraphics[height=4.0cm,width=0.35\textwidth]{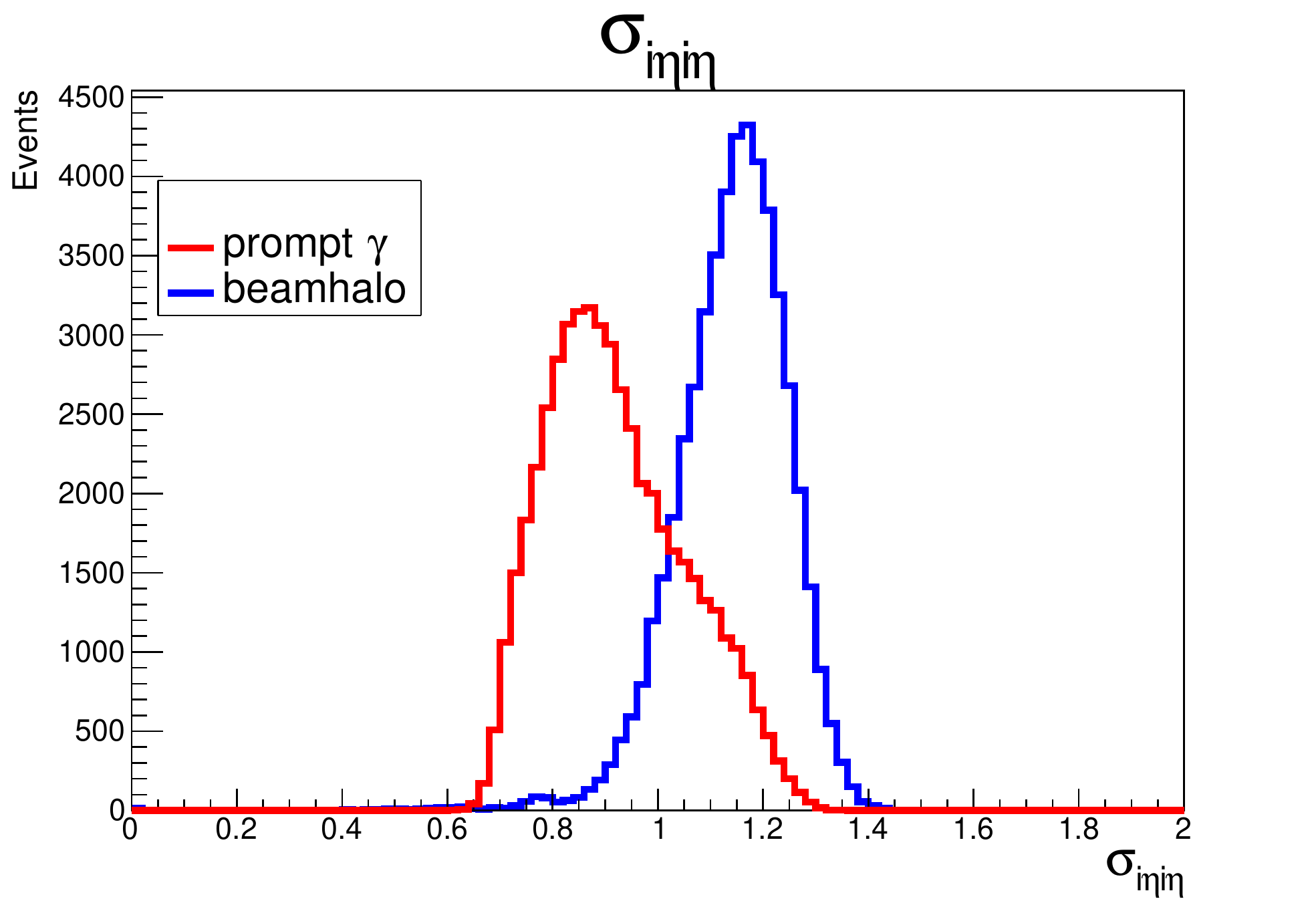}}
\subfloat[(c)]{\includegraphics[height=4.0cm, width=0.35\textwidth]{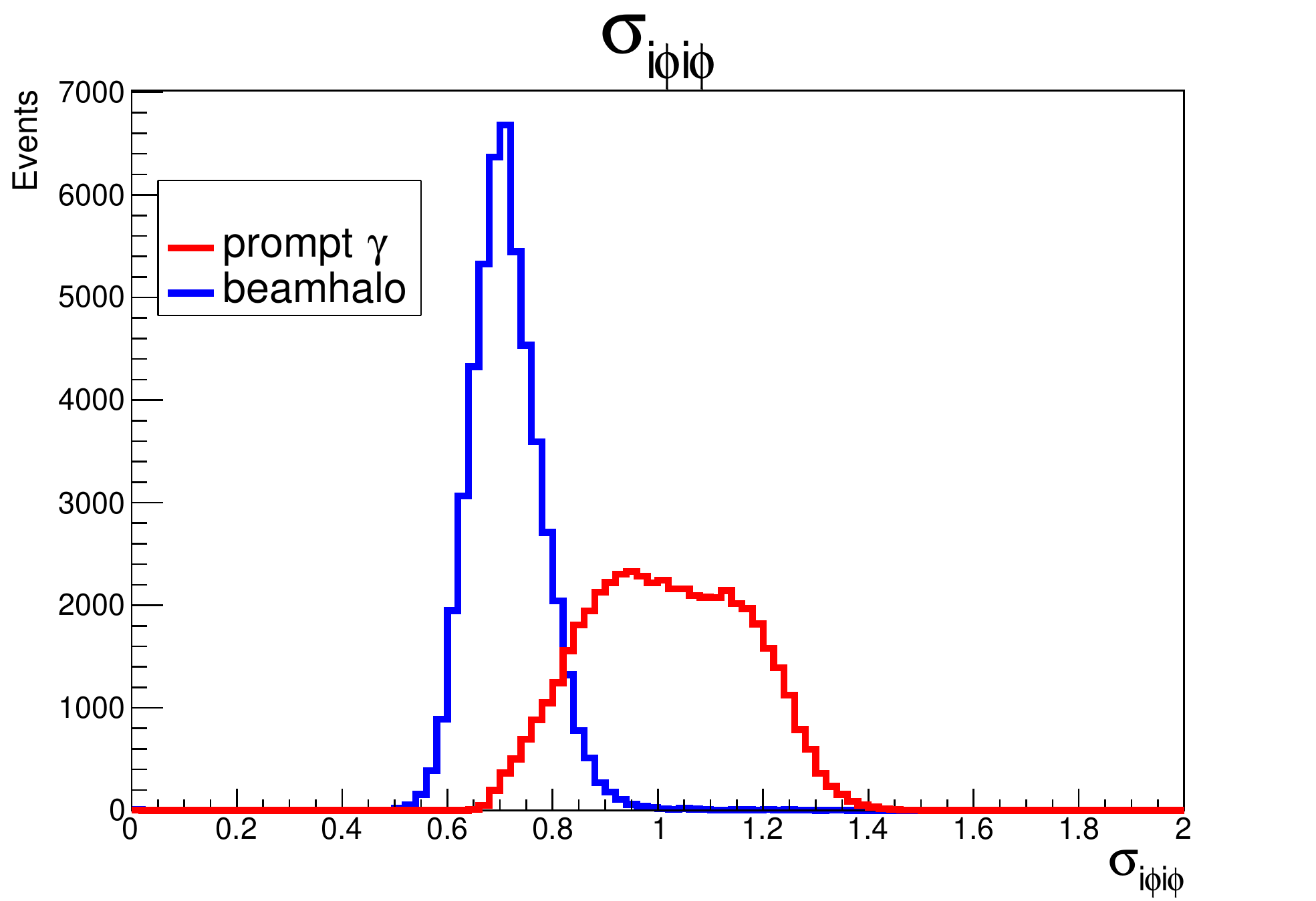}}
\caption{The distribution of the shower shape variables of 10 GeV prompt photons and beam halo photons: (a) s9/s25, (b) $\sigma_{i\eta i\eta}$, and (c) $\sigma_{i\phi i\phi}$ . }
\label{fig:halo_showershape}
\end{figure}

Samples of 10 GeV prompt photons and beam halo photons have been generated using
the above mentioned setup. These samples were first analyzed with shower shape
variables to capture the differences in the energy spread in the 
$\eta \times \phi$ space as shown in Figure~\ref{fig:all_photon}. These
variables are:
$$ s_9/s_{25}, \hspace{1.0cm} \sigma_{i\eta i\eta}, 
\hspace{1.0cm}\sigma_{i\phi i\phi}$$
where, $s_9/s_{25}$ is the ratio of the energy contained within the $3\times3$ 
matrix of crystals centered on the seed crystal to the total energy contained 
in the $5\times5$ matrix around the seed crystal and
\begin{equation}
\sigma^2_{i\eta i\eta} = \frac{\Sigma_{i=1}^{5\times5} 
w_i (i\eta_i-i\eta_{seed})^2}{\Sigma_{i=1}^{5\times5}  w_i},
\hspace{0.5cm} w_i= \mbox {max} \left(0 , 4.7+ \ln \frac{E_i}{E_{5\times5}}
\right)
\end{equation}
with $E_i$ and $i\eta_i$ being the energy and $\eta$ index of the $i^{th}$ 
crystal within the 5$\times$5 cluster and $i\eta_{seed}$ being the $\eta$ 
index of the seed crystal~\cite{CMS-PAS-EGM-10-005}.
Similarly for $\phi$ we have,
\begin{equation}
\sigma^2_{i\phi i\phi} = \frac{\Sigma_{i=1}^{5\times5} 
w_i (i\phi_i-i\phi_{seed})^2}{\Sigma_{i=1}^{5\times5}  w_i}, 
\hspace{0.5cm} w_i= \mbox {max} \left(0 , 4.7+ \ln \frac{E_i}{E_{5\times5}}
\right)
\end{equation}
With the beam halo photons having an elongated spread in $\eta$ and the prompt 
photons with a more circular spread in $\eta \times \phi$ space, these 
variables have been chosen aiming to distinguish between the two classes 
utilizing these differences. The distributions of these variables are shown 
in Figure~\ref{fig:halo_showershape} for the two samples and one can see that
both of the $\sigma^2$ variables have more separation power than the
$s_9/s_{25}$ variable. A combination of these three variables may give yet
better separation power. Thus we feed these three shower shape
variables to an MLP and obtain a background (beam halo photon) rejection 
rate to be around $71\%$ for $99\%$ signal efficiency, see 
Table~\ref{table1}. We note that such a low background rejection rate is not sufficient to perform any useful analysis.
\begin{table}[htbp]
\centering
\caption{Results for beam halo - prompt photon separation }
\label{table1}
\begin{tabular}{|l|c|c|c|}
\hline
Method used     & \begin{tabular}[c]{@{}l@{}} Number of parameters  \end{tabular} & \begin{tabular}[c]{@{}l@{}}Background rejection for \\ 99\% signal efficiency\\              (\%)\end{tabular}  & ROC AUC \\ \hline
CNN on image        & 13,199    & 99.96    & 0.9997  \\ \hline
MLP on image        & 18,732    & 99.89    & 0.9990  \\ \hline
MLP on 3 variables  & 206    & 71.31    & 0.9748  \\ \hline
\end{tabular}
\end{table}
\begin{figure}[ht]
\subfloat[(a)]{\includegraphics[ width=0.5\textwidth]{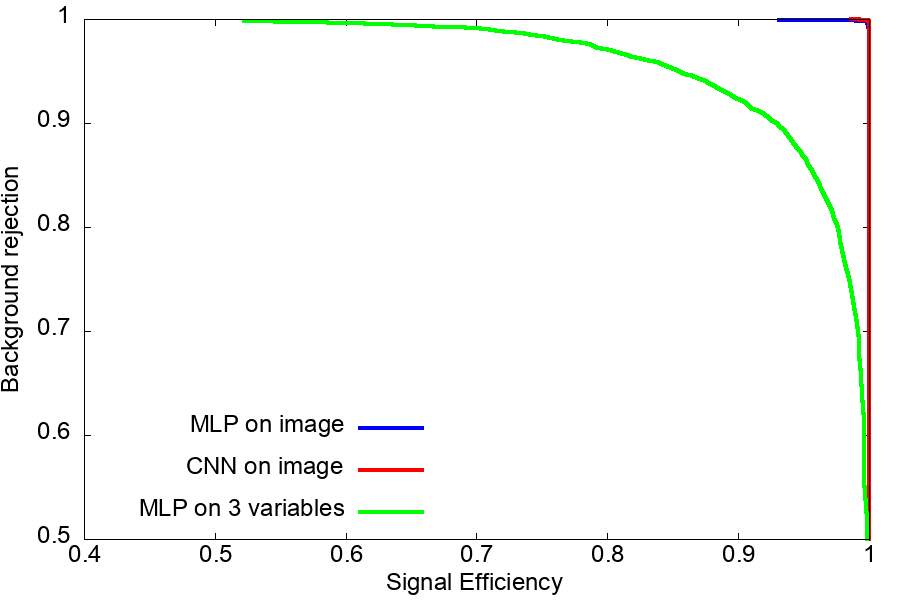}}
\hspace{0.3cm}
\subfloat[(b)]{\includegraphics[ width=0.5\textwidth]{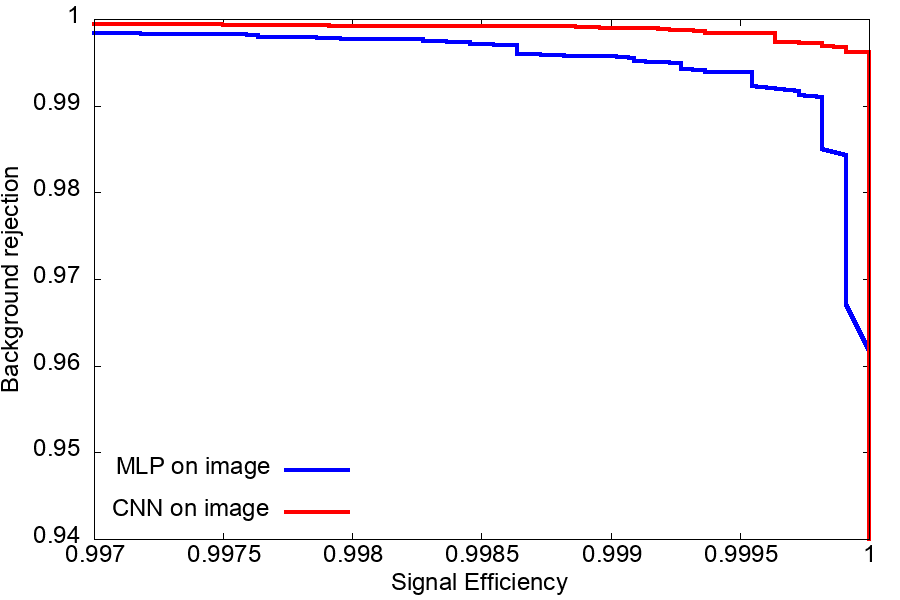}}
\caption{(a) ROCs for all methods used in separation of prompt photons and beam halo photons, (b) zoomed-in view of the ROCs which shows that the CNN performs better than the MLP.}
\label{fig:b514-halo_roc}
\end{figure}

Instead of using just three representative shower shape variables constructed
from $5\times5$ image, if we use a larger $11\times11$ image centered around the
seed crystal in the above mentioned MLP, the number of free parameters grow from
around 200 to above 18000. However, this leads to a remarkable improvement in
the background rejection rate from $71\%$ to $99.89\%$ and the method starts to
look attractive for realistic analysis. Further, if we use a CNN of the same
image size sample, the number of parameter reduces to about two thirds while the
background rejection rate improves to $99.96\%$. In other words, the false
positive rate reduces from $0.11\%$ to $0.04\%$ when we go from MLP to CNN. The
comparisons of the networks is given in Table~\ref{table1} and the
corresponding ROC shown in Figure~\ref{fig:b514-halo_roc} for completeness.


\subsection{$\pi^0-$ prompt photon separation}
\begin{figure}[]
\centering
\includegraphics[width=0.5\textwidth]{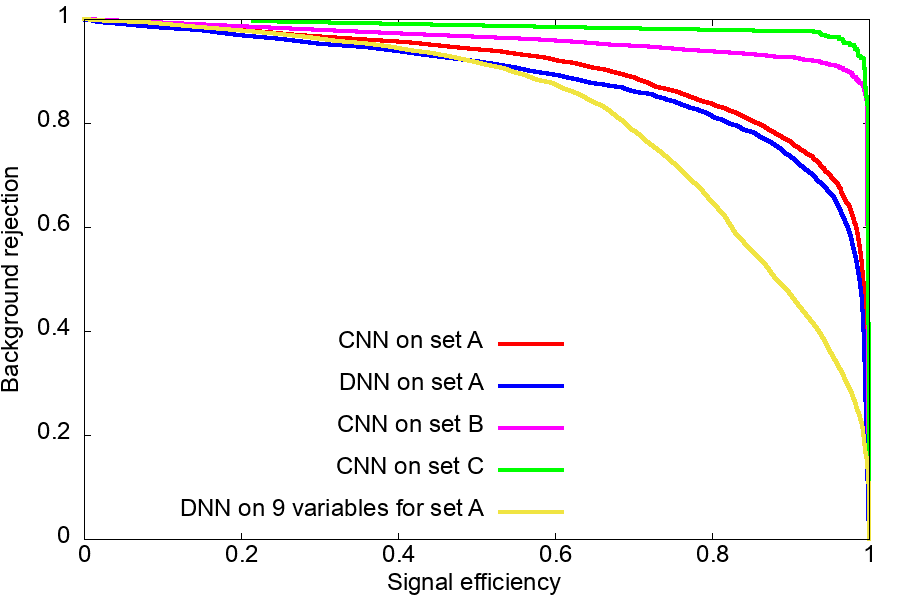}
\caption{ROCs for all the methods used for  separation of $\pi^0$s and prompt photons.}
\label{fig:ROC_pi0}
\end{figure}

In our simulation, the sample containing prompt photons has about 38\% chance of getting converted into $e^+e^-$ pairs leading to a different signature (see Figure~\ref{fig:all_photon}). The $\pi^0$ sample has two photons, and there is about 61\% chance that at least one will be converted.\\
The samples have been grouped into different sets as follows:
\begin{itemize}
    \item Set A - converted and unconverted photons vs. converted and unconverted $\pi^0$'s
    \item Set B - unconverted photons vs. converted and unconverted $\pi^0$'s
    \item Set C - unconverted photons vs. unconverted $\pi^0$'s
\end{itemize}
The CNN and the DNN have been run on these sets of 25$\times$25 images, and from the ROCs in Figure~\ref{fig:ROC_pi0}, it is evident that the CNN outperforms the DNN.\\ \\
The DNN is fed with the following set of shower shape variables centered around the maximum energy seed crystal in a 5$\times$5 cluster as the input:

    $$ s_1/s_{25}, s_4/s_{25}, s_9/s_{25}, s_{16}/s_{25}, s_{nmax}, \sigma_{i\eta i\eta}, \sigma_{i\phi i\phi}, \sigma^2_{i\eta i\phi}, r_9 $$

\begin{table}[h]
\centering
\caption{Results for $\pi^0 - \gamma$ separation}
\label{table2}
\begin{tabular}{|l|c|c|c|}
\hline
Method used     & \begin{tabular}[c]{@{}l@{}} Number of parameters  \end{tabular}   & \begin{tabular}[c]{@{}l@{}}Background rejection for\\ 90.0\% signal efficiency    (\%)\end{tabular}  & ROC AUC \\ \hline
DNN on 9 variables for set A     & 2,876      & 46.8      & 0.8196  \\ \hline
DNN on image set A               & 433,460      & 73.4      & 0.8825  \\ \hline
CNN on image set A               & 100,559      & 76.4      & 0.9030  \\ \hline
CNN on image set B               & 100,559      & 92.5      & 0.9567  \\ \hline
CNN on image set C               & 100,559      & 97.7      & 0.9848  \\ \hline

\end{tabular}
\end{table}
Along with variables defined in section~\ref{S:halo}, $s_1$ is the energy of the seed crystal, $s_4$ is the maximum energy contained within a 2$\times$2 matrix containing the seed crystal, $s_{16}$ is the maximum energy contained within a 4$\times$4 matrix containing the seed crystal, $s_{nmax}$ is $E_i/E_{seed}$ where $E_i$ is the energy of the crystal with the next highest energy adjacent to the seed crystal and $E_{seed}$ is the energy of the seed crystal, $r_9$ is the ratio of s$_9$ with the energy in the 25$\times$25 matrix centered on the seed crystal, and
\begin{equation}
    \sigma^2_{i\eta i\phi} = \frac{\Sigma_{i=1}^{5\times5} w_i (i\eta_i-i\eta_{seed})(i\phi_i-i\phi_{seed})}{\Sigma_{i=1}^{5\times5}  w_i}, \hspace{0.5cm} w_i= \mbox {max} \left(0 , 4.7+ \ln \frac{E_i}{E_{5\times5}}\right)
\end{equation}
%
\begin{figure}[ht!]
\subfloat[(a)]{\includegraphics[
width=0.35\textwidth]{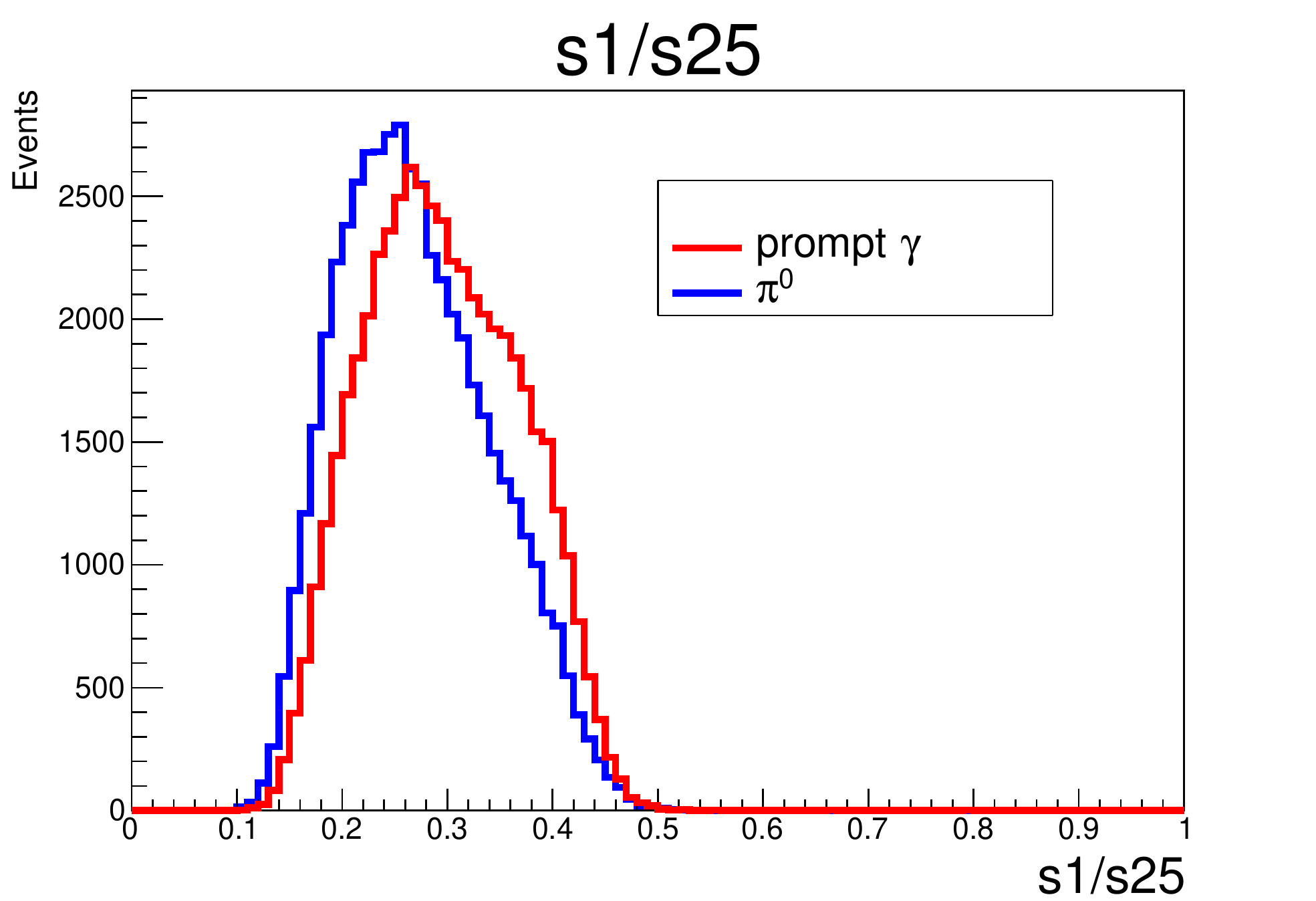}}
\subfloat[(b)]{\includegraphics[ width=0.35\textwidth]{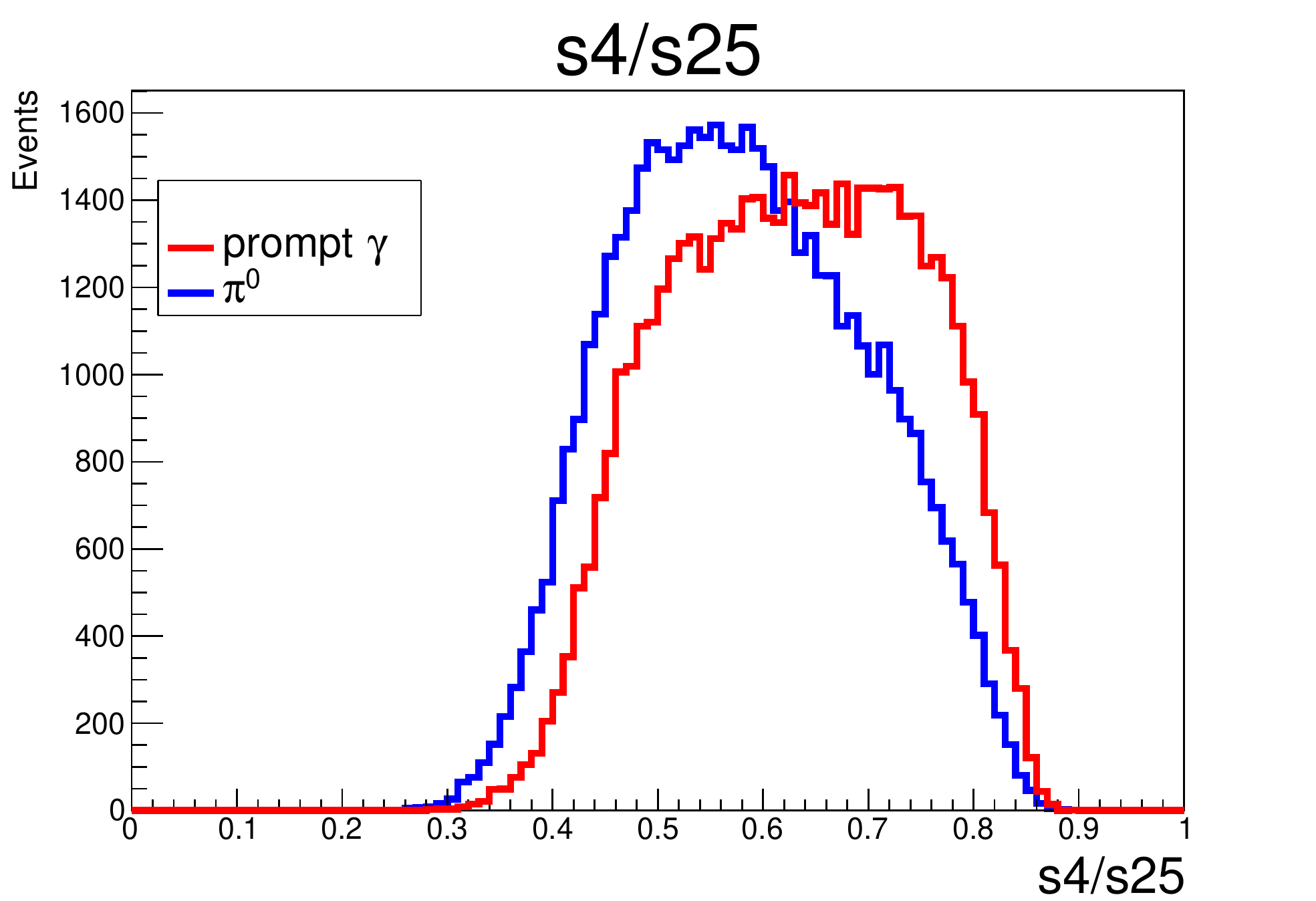}}
%
\subfloat[(c)]{\includegraphics[ width=0.35\textwidth]{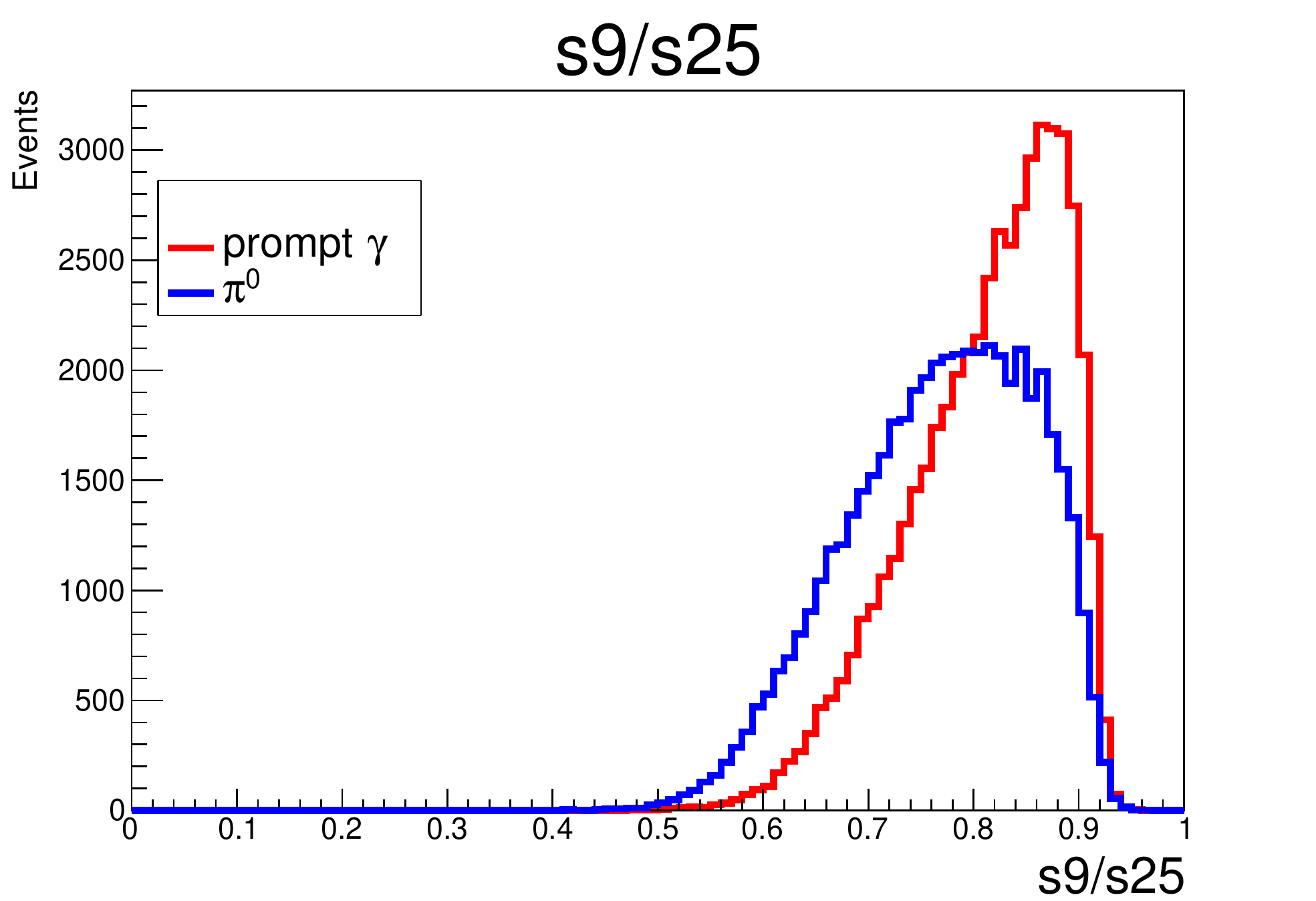}}
\vskip 0.001pt
\subfloat[(d)]{\includegraphics[ width=0.35\textwidth]{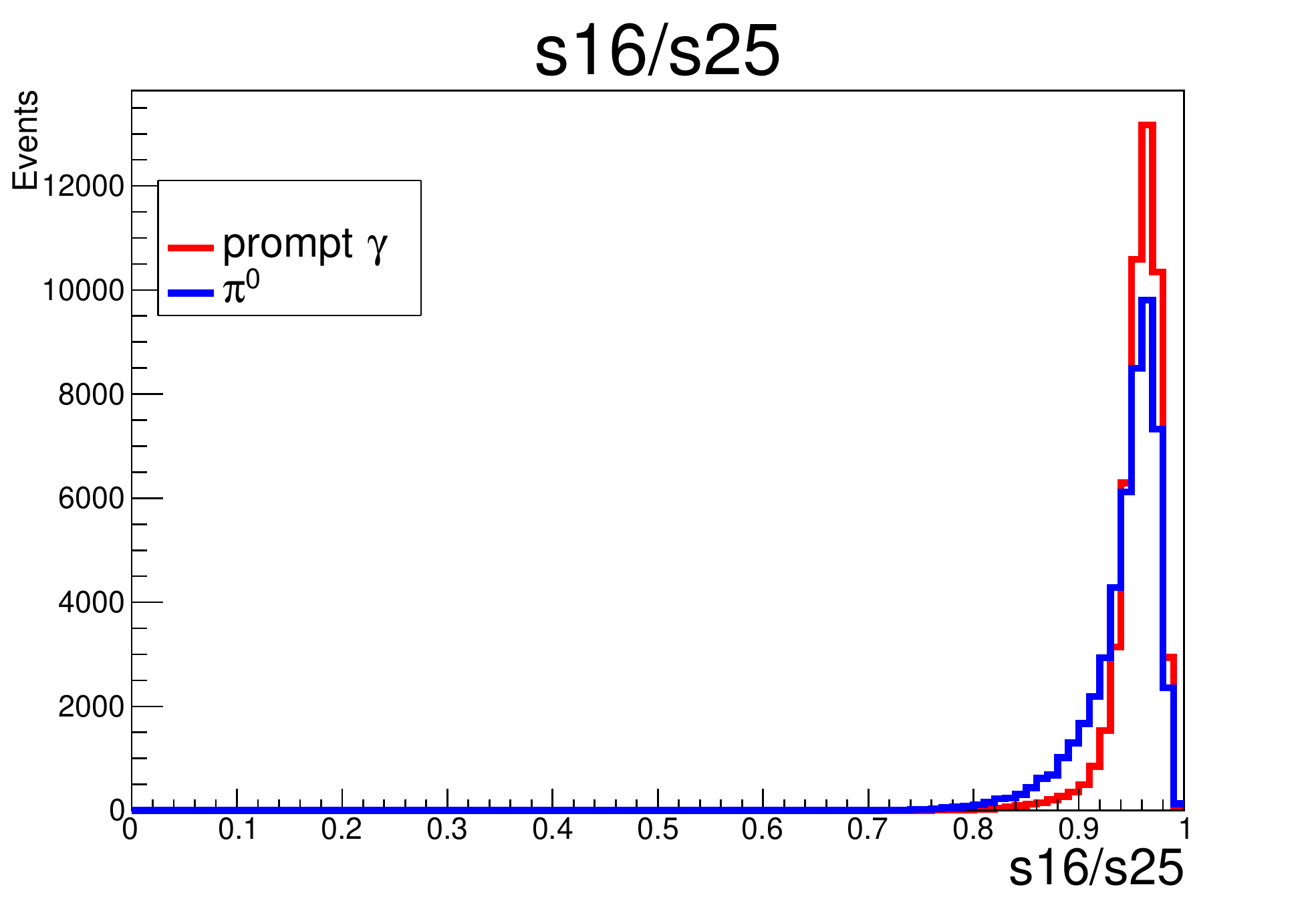}}
\subfloat[(e)]{\includegraphics[ width=0.35\textwidth]{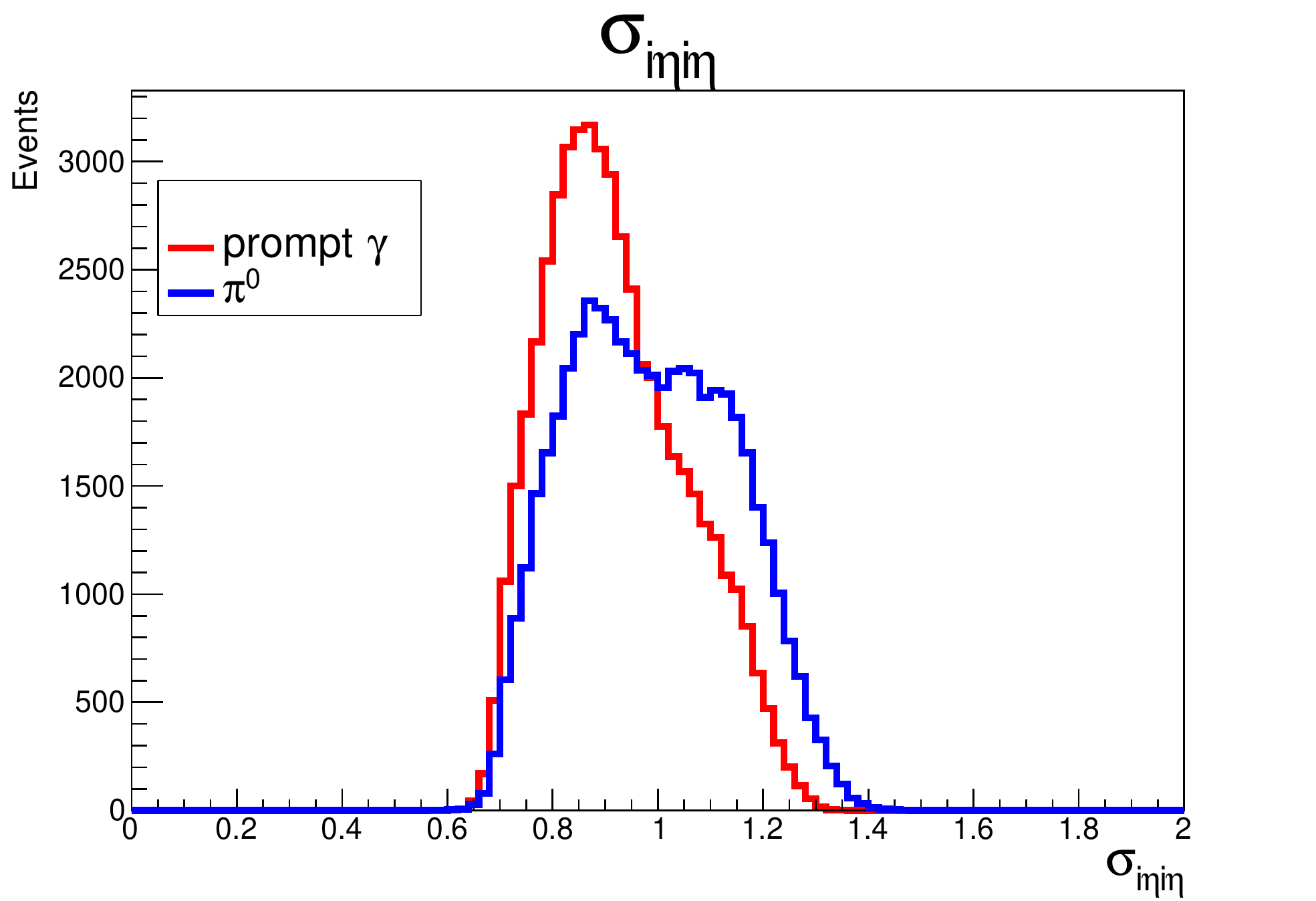}}
\subfloat[(f)]{\includegraphics[ width=0.35\textwidth]{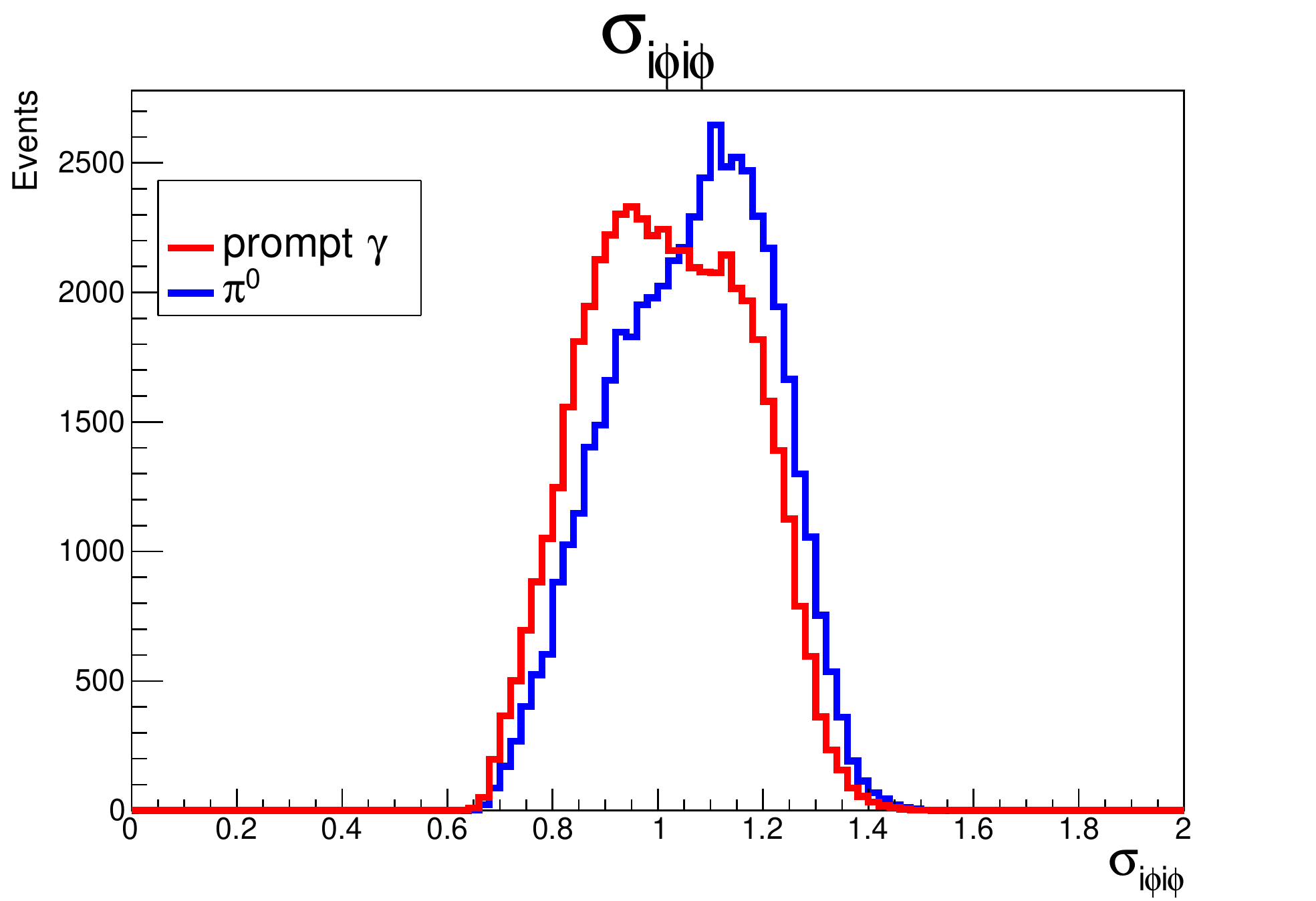}}
\vskip 0.001pt
\subfloat[(g)]{\includegraphics[ width=0.35\textwidth]{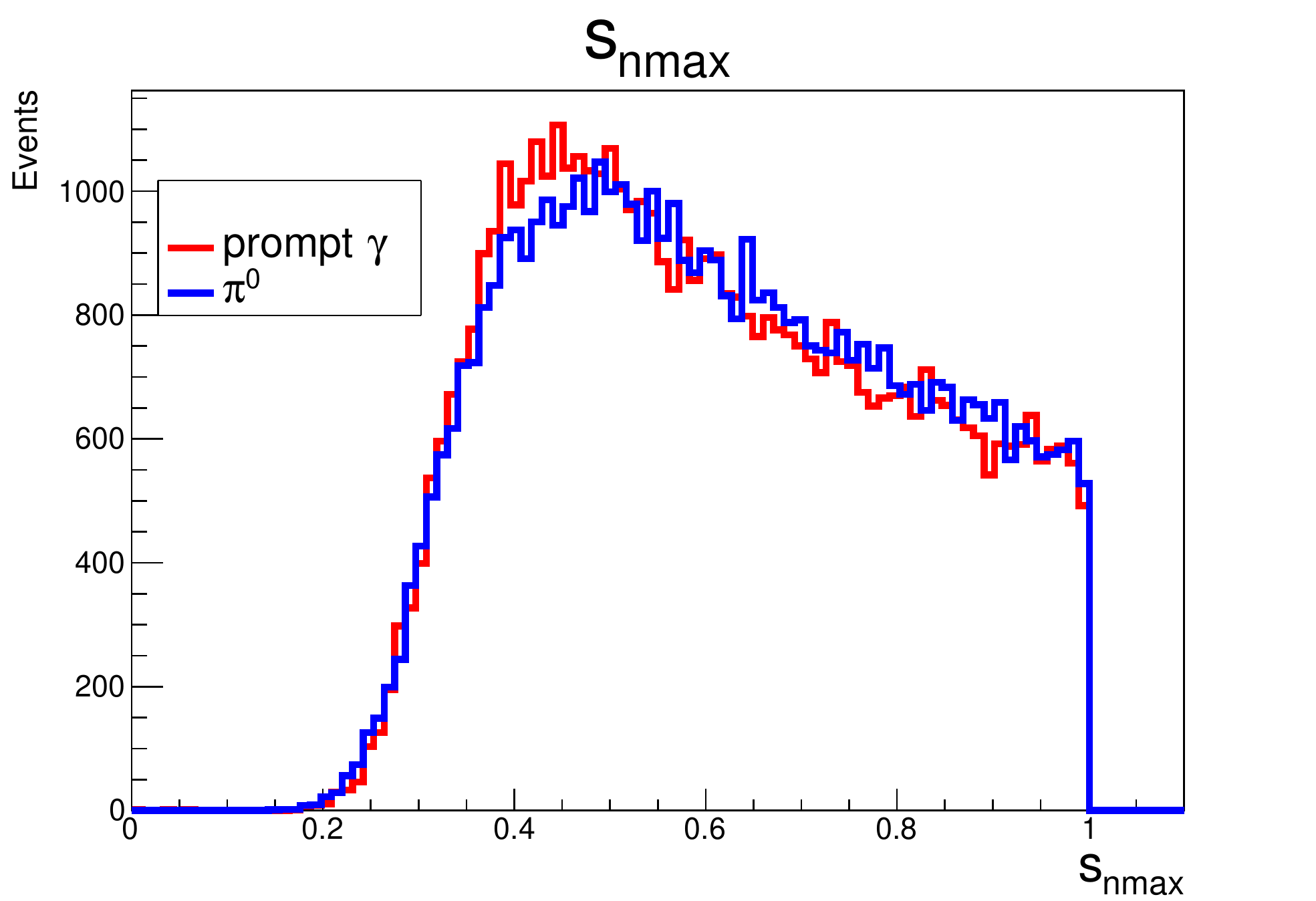}}
\subfloat[(h)]{\includegraphics[ width=0.35\textwidth]{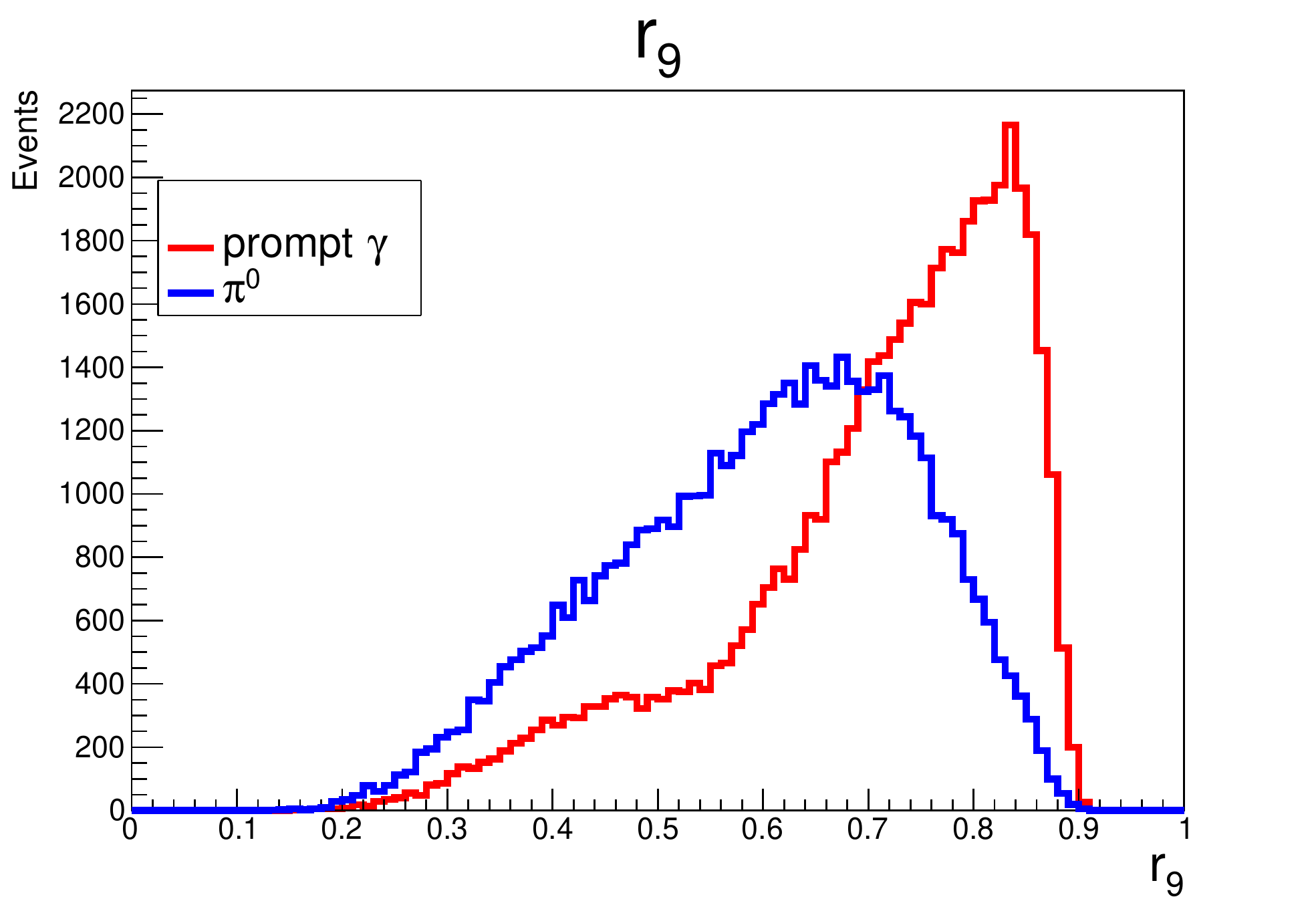}}
\subfloat[(i)]{\includegraphics[ width=0.35\textwidth]{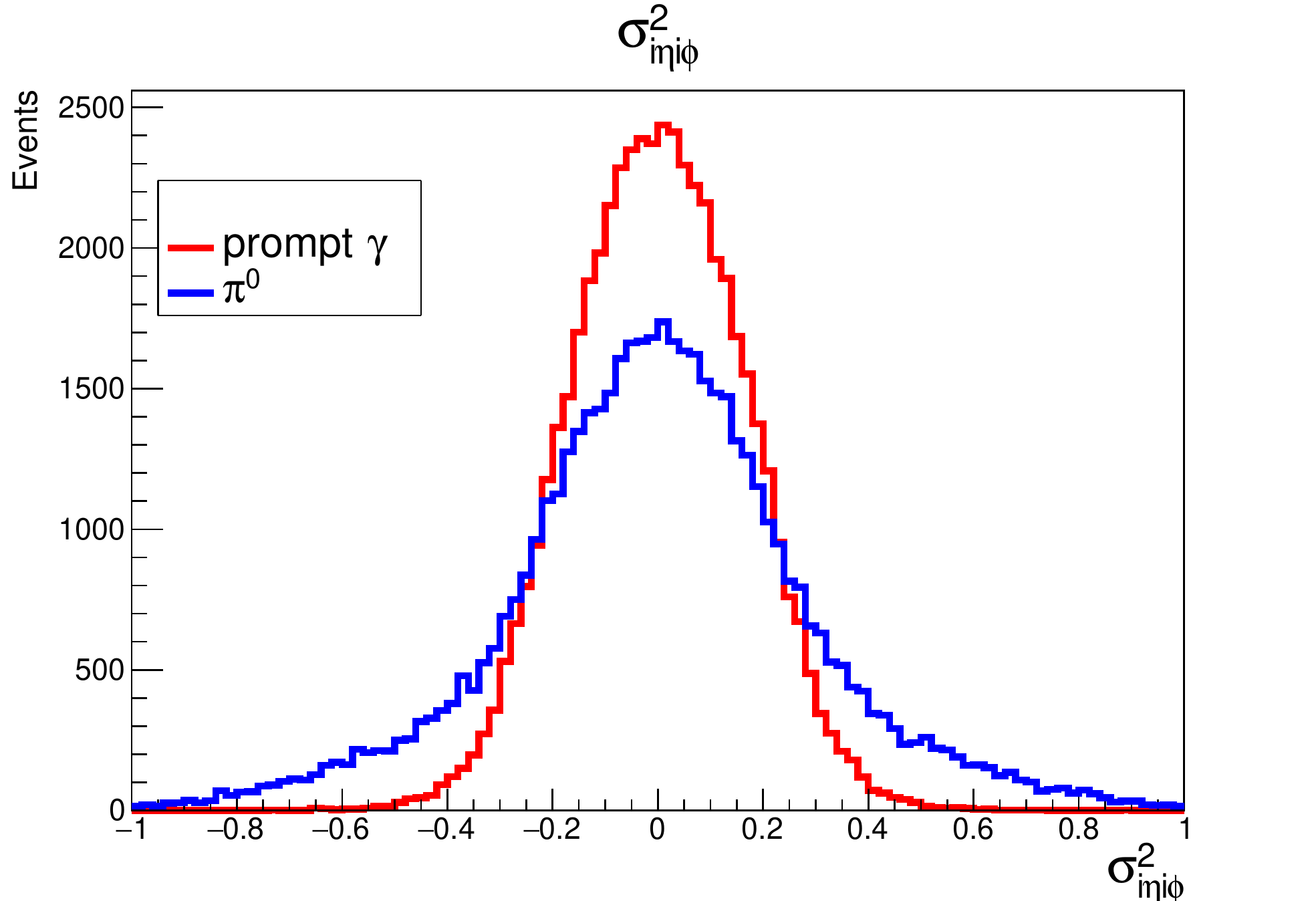}}
\caption{The distribution of the shower shape variables for set A of prompt photons and $\pi^0$'s:\\ (a) $s1/s25$, (b) $s4/s25$, (c) $s9/s25$, (d ) $s16/s25$, (e) $\sigma_{i\eta i\eta}$, (f) $\sigma_{i\phi i\phi}$, (g) $s_{nmax}$, (h) $r_9$, and (i) $\sigma_{i\eta i\phi}^2$ . }
\label{fig:showershape}
\end{figure}
The distributions of each of these variables for set A of photons and $\pi^0$'s are shown in Figure~\ref{fig:showershape}. 
These variables do not have much distinguishing power and it can be seen from the ROCs in Figure~\ref{fig:ROC_pi0} as well as from Table~\ref{table2} that the DNN trained with these variables perform the worst at 46.8$\%$ background rejection. The DNN trained on the images gives a much better background rejection of 73.4$\%$, but comes at a higher computational cost of 150x more parameters. The CNN fed with the images on the other hand gives a marginal (3$\%$) increase in background rejection but at about a quarter of the parameter cost. Although we get the best signal and background separation for Set C where both prompt photon and photons from $\pi^0$ are unconverted we would like to point out that here we have just used calorimeter information and with tracker information for both Set A and B, we would get much better background rejection numbers for them.
The background rejection, signal efficiency and the area under the curve of the ROCs are listed in Table~\ref{table2} for completeness.

\section{Conclusion}
\label{S:5}
A study has been presented using deep learning techniques for separating prompt photons from neutral pions and beam halo. Different network types have been compared on simulation data from approximated CMS detector geometry consisting of a tracker and a calorimeter.
\\ For separating photons from beam halo, CNN based on image gives the maximum background rejection of 99.96$\%$ for 99.00$\%$ signal efficiency. 
For separating neutral pions from prompt photons, CNN based on image gives the maximum background rejection of 97.7$\%$ for 90.0$\%$ signal efficiency. For both the cases, it is evident from the ROC AUCs that the CNN outperforms the MLP or DNN with the same input image as well as the MLP or DNN using topological variables.\\ For the beam halo separation, the spatial patterns of the prompt photon and beam halo photon are so distinct that a simple one layered MLP will suffice to do a high accuracy classification. The $\pi^0-\gamma$ classification proved to be a more difficult problem owing to the very similar energy deposition patterns of the $\pi^0$ decay photons and a prompt photon. The CNN still gives a good performance on unconverted photons vs. unconverted $\pi^0$'s.
These techniques of neural networks based on images are generic and can be applied to any calorimeter. 
\newpage
\acknowledgments
The authors would like to thank Dr.\ Andre Holzner for useful discussions through the course of this project and also for meticulously going through the text and providing valuable comments. The authors would also like to thank Dr.\ Ananda Dasgupta for the useful discussions through the course of this project.






\bibliographystyle{JHEP}
\bibliography{mybib.bib}
\end{document}